# Feature selection for longitudinal microarray data by adapting a pathway analysis method

Suyan Tian[1,2][§], Chi Wang[3], Howard H. Chang[4]

[1] Division of Clinical Epidemiology, The First Hospital of Jilin University, 71Xinmin Street, Changchun, Jilin, China, 130021
[2] School of Mathematics, Jilin University, 2699 Qianjin Street, Changchun, Jilin, China, 130012
[3] Department of Biostatistics, Markey Cancer Center, The University of Kentucky, 800 Rose St., Lexington, KY, 40536, 1518 Clifton Road NE, Atlanta, GA, 30322
[4] Department of Biostatistics and Bioinformatics, Rollins School of Public Health, Emory University

[§]Corresponding authors

Email addresses:

    ST:    stian@rockefeller.edu
    CW:    chi.wang@uky.edu
    HHC:   howard.chang@emory.edu

# Abstract

**Introduction**

Feature selection and gene set analysis are of increasing interest in bioinformatics. While these two approaches have been developed for different purposes, we describe how some gene set analysis methods can be used to conduct feature selection. Here we adapt the gene set analysis method, significance analysis of microarray gene set reduction (SAMGSR), for feature selection, and propose two extensions – simple SAMGSR and two-level SAMGSR to identify relevant features for longitudinal microarray data.

**Results and Discussion**

When applied to a real-world application, both simple and two-level SAMGSR work comparably well. Using simulated data, we further demonstrate that both SAMGSR extensions have the ability to identify the true relevant genes. If the relevant genes are not highly correlated with the irrelevant ones, the final models given by the two SAMGSR extensions are parsimonious as well.

**Conclusions**

By adapting SAMGSR for feature selection and applying the proposed algorithms on a longitudinal gene expression dataset, we demonstrate that a gene set analysis method can be used for the purpose of feature selection. We believe this work paves the way for more research to bridge feature selection and gene set analysis with the development of novel algorithms.

# Introduction

Feature selection and gene set analysis are currently two tools in bioinformatics being employed with increasing frequency. While gene set analysis aims to identify relevant gene sets associated with a phenotype of interest, feature selection focuses on the identification of relevant individual genes. Previous studies have demonstrated that the performance of a feature selection algorithm, when incorporating biological information or co-expression patterns conveyed by those gene sets, is usually superior [1–4].

Alternatively, a gene set analysis algorithm can be adopted directly to identify relevant features. For instance, a novel gene set analysis method called significance analysis of microarray - gene set reduction analysis (SAMGSR) was proposed by [5], aiming at a further reduction of selected gene set into respective core subsets. The concise core subsets in SAMGSR facilitate our understanding of biological mechanisms underlying the phenotype of interest. The reduction process can bee viewed as feature selection, which motivated us to adopt SAMGSR algorithm for feature selection. By applying SAMGSR to a real-world microarray data, we found that the resultant gene signature can distinguish multiple sclerosis (MS) patients apart from their controls with a good accuracy [6].

Because biological systems or processes are dynamic, researchers have been interested in investigating gene expression patterns longitudinally, aiming at capturing biologically meaningful dynamical changes. With the fast evolution of microarray technology, longitudinal microarray experiments have become affordable and increasingly common in different fields. Typically, regarding to data analysis, such data have been tackled by stratifying the data into separate time points and analyzing them separately. This approach can be inefficient because it ignores the highly dependent structure of longitudinal data. Also, changes in gene expression pattern over time may not be detected by considering the results from marginal analysis at each time point individually, leading to loss in statistical power [7–9]. Consequently, most of the feature selection methods had been originally developed for a cross-sectional study, where the outcome is measured at a single time point, become ineffective for longitudinal 'omics' data [9].

Many methods have been proposed for longitudinal gene expression data either by adapting traditional strategies used in longitudinal data analysis or by developing complex statistical models. For instance, Storey et al [7] proposed a method designed to identify differentially expressed gene over time between different phenotypes. This method utilized spline-based models to estimate expression-versus-time curves for genes individually and a specific software program called EDGE [10] has been developed for implementation. However, this algorithm belongs to the simplest feature selection category –filter [11]. A filter method tends to select all highly correlated features into the final model, thus resulting in an inferior model in terms of parsimony. In their work [7], substantially a large number of genes were deemed as significant by EDGE, but not by t-test and SAM-test [12]. Similar phenomenon was observed in one recent study [13] where EDGE was applied to characterize transcriptomic difference after severe trauma and burn injury compared to healthy subjects, with more than 80% of genes showing significant differences between the two groups. Regarding to this unexpected "genomic storm", we postulate the true positive rate may suffer.

In this paper, we propose two extensions to SAMGSR to conduct feature selection for longitudinal microarray data. First, we extend SAMGSR by applying its reduction to core subset part twice hierarchically. Specifically, genes in selected gene sets are reduced to a core gene subset first and then time points are selected subsequently. We denote this extension the two-level SAMGSR. Second, we view one specific gene over time points as a gene set. SAMGS is used to identify significant genes that experience change in expression values, and then SAMGSR selects at exact which time point/s correspond to significant difference in gene expression. This extension is named as simple SAMGSR.

## Methods and Materials

### Experimental Data

Data of the real-world application were retrieved from GEO (accession number GSE36809). It was hybridized on Affymetrix HGU133 plus2 chips. In this cohort, there were 167 severe blunt trauma patients. Because our objective is to select genes that can distinguish between trauma patients with complications apart from those without, only patients with uncomplicated recovery and patients with complicated recovery were considered. The definitions for complicated and uncomplicated recovery are given in the original study [13]. Briefly based on the duration of recovery, uncomplicated recovery represents recovery within 5 days while complicated one is the recovery after 14 days, no recovery by 28 days, or death.

In this study, we only included those patients without complication with five time points, and patients with complication with seven time points. Moreover because for patients without complication their longest follow-up is 14 days, the data for patients with complication were truncated at 14 days. The time points we considered here in the training set were days 1/2, 1, 4, 7 and 14. Lastly, we used the rest of complicated and uncomplicated patients as a test set to validate both algorithms. Correspondingly, the time points considered in the test data were days 1/2, 1, 4, and 7.

### Pre-processing Procedures

Since different pre-processing procedures may impact analysis, we decided to download expression values ($log_2$ transformed) of the experimental data from the GEO database (http://www.ncbi.nlm.nih.gov/geo/). We did not conduct any pre-processing.

### Pathway information

The gene sets were downloaded from the **Molecular Signatures Database** (MSigDB) [14]. In this study, we only considered C2 category in this database, which includes gene sets from curated pathways databases such as KEGG [15] and those manually curated from the literature on gene expression. Among 17,935 unique genes in the data set, 16,121 are annotated by C2 category.

### Statistical Methods

Here, we present a brief introduction to SAMGSR, and discuss in detail about our extensions for the purpose of feature selection on longitudinal gene expression data.

### *SAMGSR*

SAMGSR is an extension of SAMGS [16] that provides further reduction of significant gene sets into respective core subsets. In SAMGSR with an approximately 90% of reduction in the size of selected genes, such that predictive performance is improved and biological patterns become more obvious.

Briefly, SAMGSR consists of two major steps [5]. First step is essentially the SAMGS process. In this step, an SAM-GS statistic, the $L_2$ norm of the SAM statistics [12] for all genes within that gene set, is calculated. Using a permutation test by perturbing phenotype labels to calculate p-values, statistical significance of a gene set is determined. Second, on those significant gene sets SAMGSR orders genes in gene set *j* based on the magnitude of its SAM statistic and then partitions gradually an entire get set

S into two subsets: the reduced subset $R_k$ which includes the first k genes with largest SAM statistic, and the residual one $\overline{R}_k$ being the complement of $R_k$ for k=1,…, |S-1|. Here S is the size of the specific gene set. Let $c_k$ be the SAMGS p-value of the residual subset $\overline{R}_k$. The optimal size of reduced set $R_k$ is the least k so that $c_k$ is larger than a pre-specified cut-off, e.g., 0.05. Conceptually, the significance of a gene within a gene set is determined by its contribution to the SAMGS statistic. It implies that if in a gene set $|SAM_i| > |SAM_j|$ for genes *i* and *j*, then gene *j* is likely to involve in the reduce subset only when gene *i* does.

*Modifications to SAMGSR*
We propose two extensions to SAMGSR – simple SAMGSR and two-level SAMGSR to carry out feature selection for longitudinal microarray data.

**1) Two-level SAMGSR**
In this extension, the reduction procedure of SAMGSR was applied in a hierarchical way in two levels: a lower level on gene sets and an upper level on the time points. First, the reduction procedure of SAMGSR was applied to identify the reduced core subsets corresponding to the selected gene sets by SAMGS. Then upon the union of genes involved in those core subsets, the reduction procedure was applied again to determine at which time points those genes differ significantly. The main advantage of the two-level SAMGSR over simple SAMGSR is that it accounts for biological information contained in pathways (e.g., co-expression pattern). Specifically, at the gene-set level, the following SAMGS statistic is defined,

$$SAMGS_{GS_k} = \sum_{i=1}^{|GS_k|} \sum_{j=1}^{t_i} d_{ij}^2, \quad d_{ij} = (\overline{x}_d(ij) - \overline{x}_c(ij))/(s(ij) + s_{0j})$$

where $d_{ij}$ is the SAM statistic [12] of gene *i* (i=1,…, $|GS_k|$) at time point *j* (j=1,…, $t_i$) and $|GS_k|$ is the size of gene set $GS_k$, here k=1,…,K, and K is the total number of gene sets. This SAMGS statistic is the $L_2$ norm of SAM statistics accumulating over all genes inside one gene set across all time points. Then in the reduction step, SAMGS is calculated sequentially to a series of subsets (for the size of 1,…, $|GS_k|$-1) of a significant gene set, aiming at identifying a core subset that makes essential contribute to the statistical significance of this gene set. This reduction part is firstly implemented upon genes, where a combination of a small number of genes with large values of $\sum_{j=1}^{t_i} d_{ij}^2$ is chosen. Then the algorithm moves to the level of time points, with the objective of determining which combination of time points contributes substantially to the importance of a gene, where $d_{ij}$ is evaluated.

**2) Simple SAMGSR**
In this extension, we consider each gene as a gene set. Namely, a gene set consists of one specific gene over different time points. Our rational is that one gene's expression values for the same individual over time are correlated, mimicking a gene set. First we applied SAMGSR to select the relevant genes and determine exactly at which time points the expression values of one specific gene differ between two phenotypes. A drawback of simple SAMGSR is it does not incorporate valuable biological information contained in pathways, which provides knowledge on how genes function in consensus to impact on biology processes. Figure 1 elucidates both simple and two-level SAMGSR algorithms. Then upon the identified genes by SAMGSR extensions, we fit support vector machine (SVM) models to compute their respective performance statistics.

*Statistical Metrics*

As in the previous study [17], we used four metrics - Belief Confusion Metric (BCM), Area Under the Precision-Recall Curve (AUPR), Generalized Brier Score (GBS), and misclassified error rate - to evaluate the performance of a classifier. The references therein described those metrics in details. Briefly, they all range from 0 to 1. For the first two metrics the closer to 1 the better a classifier is, whereas the direction is opposite for the last two metrics. For longitudinal data these metrics were calculated at each time point.

**Statistical language and packages**

Statistical analysis was conducted in the R language version 3.1.2 (www.r-project.org), and R codes for SAMGSR were downloaded from (http://www.ualberta.ca/~yyasui/homepage.html). The Venn-diagram plot was made with the aid of an online bioinformatics tool (http://bioinformatics.psb.ugent.be/webtools/Venn/). R codes of simple and two-level SAMGSR algorithms are available upon request.

# Results and Discussion

**Real data**

Traumatic injury with subsequent infection was a common cause of death in ancient time. Even today massive injury such as combat wound remains life-threatening [18, 19]. A large clinical study that examined the genome-wide expression patterns of blood leukocytes in the immediate post-injury period was carried out recently [13]. One primary objective of that study was to explore if there were different patterns of gene expression associated with two extremes of clinical recovery: uncomplicated and complicated. The authors used the EDGE algorithm and identified 2,391 differentially expressed genes (DEGs) at any one point over the time course. We used the longitudinal gene expression data collected specifically for this objective to evaluate our proposed extensions to SAMGSR, aiming at further reducing the number of selected genes.

*Results of simple and two-level SAMGSR*

Based on the performance statistics in Table 1, both simple and two-level SAMGSR work comparably well on discriminating complicated samples from uncomplicated ones at each time point. Nevertheless, incorporating extra information inside gene sets results in the two-level SAMGSR having superior overall performance across all time points, despite its additional cost on computing time and complexity. Notably, those canonically curated databases on pathways/gene sets are more comprehensive for well-studied diseases like cancers. With few work has investigated on traumatic injury using gene expression profiles, the implementation of gene set level on the dataset might not reflect the merit of two-level SAMGSR fully.

The numbers of selected genes at these 5 time points vary, but there exists a clear pattern with the least number at the middle time points and largest numbers at the tails. In Figures 2 and 3, Venn-diagrams illustrate how the selected genes by either SAMGSR extension at different time points overlap (Fig. 2), and subgroup sample mean versus time plots for those 5 genes significant across all 5 time points between patients with and without complications by both methods present a variety of change patterns over time (Fig. 3). There are two contradicted paradigms on how the host response to severe trauma: one is that complications were associated with multiple inflammatory events which may subsequently cause a secondary genomic response[20, 21]. The other is that complication results from simultaneous and rapid induction of innate and suppression of adaptive immunity genes [13]. (Figure 4 of the original paper [13] presents the graphical illustration of these two paradigms.) Nevertheless, our analysis result (Fig. 3), i.e., the mixed change patterns across time points for patients with complication versus those without provides no clear evidence on either of these two paradigms. Further investigation is warranted.

Moreover, we observe that in the 94- unique selected longitudinal gene signature by two-level SAMGSR, there is a substantial proportion of overlap at all time points (24/94). So is the 97- unique selected gene signature by simple SAMGSR (Fig. 2). In contrast, the number of genes being significant only at one specific time point is one half of this number. This sheds some evidence on that our SAMGSR extensions are capable of identifying genes that present mild but concordant difference across time points between two different phenotypes.

*Comparisons with other algorithms*
In the original work [13], 2,391 DEGs at any one point over the time course were identified using EDGE. In contrast, the maximum overall number of identified genes is less than one hundred by our SAMGSR extensions. The superiority of SAMGSR in terms of model parsimony is clear.

Furthermore, using the samples being discharged early as a test set, we evaluated the performance of the two SAMGSR extensions versus other relevant algorithms: SAMGSR, penalized SVM, and LASSO [22] at each time point. The results are also presented in Table 1. Compared to other algorithms, SAMGSR algorithms show superior performance in terms of overall model parsimony, e.g., two-level SAMGSR identifies 94 unique genes while LASSO selects 147 genes. Therefore, consideration on the biology interactions between genes involved in pathways, maybe as well as the correlations of one gene's expression over time longitudinally, might increase the capacity of excluding redundant and irrelevant features, without lowering predictive performance.

**Synthesized data**

In order to investigate the properties of both SAMGSR extensions, we used observed expression values from the injury application to design two sets of simulations.

First, we chose two causal genes – F13A1 and GSTM1 and then randomly selected 998 genes from the data serving as noise. Denote the expression value of gene $i$ (F13A1 or GSTM1) at time $j$ (j=1,…, 5) as $X_{i,j}$, the probability of an injury with complication was decided using the following logit function,

$$\log it_{c/u} = 0.18 X_{F13A1.1} + 0.57 X_{F13A1.2} + 0.29 X_{F13A1.3} + 0.41 X_{F13A1.4} + 1.02 X_{GSTM1.3}$$

In this set of simulations, we considered one gene whose significance arises from its moderate joint contribution over time and the other whose association with the outcome is large at one specific time point.

In the second simulation, we chose two genes – COX4I2 and RP9 as the relevant genes. The logit function was,

$$\log it_{c/u} = 0.56 X_{COX4I2.1} - 0.91 X_{RP9.5}$$

For both simulation settings, 50 replicates were created. The results are tabulated in Table 2B.

Interestingly, the simple SAMGSR shows no inferiority to two-level SAMGSR in both correctly selecting relevant genes and obtaining model parsimony. Regarding model parsimony, the inferiority of two-level SAMGSR might be explained by at the pathway level, a relevant gene would be involved in many gene sets. Then as a result, the number of highly correlated genes with the relevant ones might increase, with such genes end up in the final model the parsimony of two-level SAMGSR naturally suffers.

Furthermore, although in the second simulation the number of relevant time points is less than that in the first one, the number of selected genes by both algorithms is dramatically larger in the second simulation. This might be because the relevant genes in the second simulation were highly correlated with other genes compared to the first simulation. The highly correlated structure between relevant features and irrelevant ones produces a large amount of redundant features that both SAMGSR extensions especially two-level one cannot exclude. To our best knowledge, however, many feature selection algorithms, especially those based on filtering, may suffer more from this drawback. As illustrated by our previous work [23, 24], an additional filtering using a relevant algorithm such as bagging [25] may alleviate this problem.

## Conclusions

In the injury application, only patients with complications and five measurements, and patients without complications and with seven time points were included. With such restriction, patients discharged from the hospital earlier (thus having later measurements missing) were deleted, thus it might lead to including patients with more severe conditions. Similar to SAMGSR, our proposed extensions can handle missing values without any difficulty. Thus the proposed algorithms do not require this restriction, but we imposed this restriction to simplify our data analysis and to use partial data of those early-discharged patients as the test data. Since the characteristics of patients in the training set and the test set may be quite different, as expected the predictive errors were very high. However, the superiority of SAMGSR extensions in terms of model parsimony is apparent.

Using a real-world microarray dataset, we demonstrated that both SAMGSR extensions account for the correlated structure of one expression's profiles over time, thus tend to identify genes with longitudinal aggregated effects while their effect size at individual time points may be insignificant. Such genes would be missed when a regularization method was applied at individual time point level. This explains mainly why both extensions outperform in terms of model parsimony. Notably, although evaluation on individual time points using those statistical metrics might be unfair for both SAMGSR extensions given their tendency to identify those genes that are insignificant at isolated points but significant jointly over time, their predictive performance is comparable to other novel feature selection algorithms.

The curated pathways in major databases such as KEGG and GO tend to be enriched in the most prevalent disease－cancers. These genes are well studied compared to those genes sorely associated with diseases other than cancers, potentially introducing biases into a pathway-based algorithm. One solution is to consider a statistical method to construct a complete set of gene subnets for all genes under consideration e.g., [26]. Future work on how to construct such gene sets for longitudinal microarray data is needed, particularly to determine whether gene sets are stable or dynamic over time.

In this article, we adapted SAMGSR for feature selection for longitudinal gene expression profiles and demonstrated the application of gene set analysis methods for feature selection. Many gene set analysis methods, such as [27], may also be applied directly or modified correspondingly to identify real 'driving' features which characterize the phenotype of interest well.

## Abbreviations

NSCLC: non-small cell lung cancer; AC: adenocarcinoma; SCC: squamous cell carcinoma; SAMGSR: significance analysis of microarray gene set reduction; SAM: significance analysis of microarray GEO: Gene Expression Omnibus; MSigDB: Molecular Signatures Database; SVM: support vector machine; BCM: Belief Confusion Metric; AUPR: Area Under the Precision-Recall Curve; GBS: Generalized Brier Score.

## Competing interests

No competing interests have been declared.

## Authors' Contributions

Conceived and designed the study: ST. Analyzed the data: ST CW. Interpreted data analysis and results: ST HHC CW. Wrote the paper: ST HHC. All authors reviewed and approved the final manuscript.

## Acknowledgements

This study was supported by fund No. 31401123 from the Natural Science Foundation of China.

# Tables

**Table 1. Performance of SAMGSR extensions and comparison with other relevant algorithms using injury data**

|  | Training set | | | | | Test set | | | |
|---|---|---|---|---|---|---|---|---|---|
|  | Day 1/2 | Day 1 | Day 4 | Day 7 | Day 14 | Day 1/2 | Day 1 | Day 4 | Day 7 |
| A. Using two-level SAMGSR (94-gene signature, cutoff for $c_k$ = 0.2)[1] | | | | | | | | | |
| # of genes | 63 | 55 | 49 | 63 | 70 | 63 | 55 | 49 | 63 |
| Error (%) | 0 | 0 | 2.33 | 0 | 0 | 43.42 | 38.16 | 39.47 | 51.32 |
| GBS | 0.055 | 0.055 | 0.063 | 0.061 | 0.042 | 0.298 | 0.272 | 0.240 | 0.288 |
| BCM | 0.777 | 0.781 | 0.764 | 0.764 | 0.814 | 0.491 | 0.534 | 0.560 | 0.495 |
| AUPR | 0.992 | 0.992 | 0.992 | 0.992 | 0.992 | 0.494 | 0.551 | 0.594 | 0.527 |
| B. Using simple SAMGSR (97-gene signature, cutoff for $c_k$ = 0.2)[1] | | | | | | | | | |
| # of genes | 69 | 53 | 45 | 77 | 71 | 69 | 53 | 45 | 77 |
| Error (%) | 0 | 0 | 2.33 | 0 | 0 | 40.79 | 53.95 | 51.32 | 44.74 |
| GBS | 0.061 | 0.055 | 0.074 | 0.057 | 0.046 | 0.262 | 0.309 | 0.307 | 0.257 |
| BCM | 0.755 | 0.779 | 0.747 | 0.780 | 0.805 | 0.499 | 0.513 | 0.498 | 0.534 |
| AUPR | 0.992 | 0.992 | 0.991 | 0.992 | 0.992 | 0.503 | 0.521 | 0.514 | 0.572 |
| C. Using SAMGSR at each time point * (the size of signature >1000, cutoff for $c_k$ = 0.1)[1] | | | | | | | | | |
| # of genes | 360 | 30 | 61 | 42 | 782 | 360 | 30 | 61 | 42 |
| Error (%) | 0 | 6.98 | 0 | 6.98 | 0 | 47.36 | 55.26 | 42.11 | 47.37 |
| GBS | 0.066 | 0.085 | 0.051 | 0.074 | 0.055 | 0.264 | 0.295 | 0.266 | 0.296 |
| BCM | 0.745 | 0.738 | 0.792 | 0.760 | 0.775 | 0.491 | 0.486 | 0.515 | 0.492 |
| AUPR | 0.992 | 0.981 | 0.992 | 0.989 | 0.992 | 0.490 | 0.482 | 0.529 | 0.512 |
| D. Using penalized SVM at each time point (the size of signature >1000, $L_1$ penalty) | | | | | | | | | |
| # of genes | 30 | 29 | 4246 | 6223 | 1030 | 30 | 29 | 4246 | 6223 |
| Error (%) | 0 | 0 | 0 | 0 | 0 | 39.47 | 46.05 | 39.47 | 47.37 |
| GBS | 0.078 | 0.083 | 0.071 | 0.072 | 0.068 | 0.256 | 0.280 | 0.242 | 0.237 |
| BCM | 0.725 | 0.716 | 0.735 | 0.732 | 0.741 | 0.516 | 0.503 | 0.512 | 0.524 |
| AUPR | 0.992 | 0.992 | 0.992 | 0.992 | 0.992 | 0.544 | 0.509 | 0.537 | 0.578 |
| E. Using LASSO at each time point (147-gene signature) | | | | | | | | | |
| # of genes | 33 | 29 | 37 | 28 | 26 | 33 | 29 | 37 | 28 |
| Error (%) | 0 | 0 | 0 | 0 | 0 | 38.16 | 44.74 | 43.42 | 36.84 |
| GBS | $<10^{-4}$ | $<10^{-4}$ | $<10^{-4}$ | $<10^{-4}$ | $2\times10^{-4}$ | 0.311 | 0.300 | 0.332 | 0.304 |
| BCM | 0.993 | 0.992 | 0.993 | 0.992 | 0.991 | 0.533 | 0.563 | 0.542 | 0.544 |
| AUPR | 0.992 | 0.992 | 0.992 | 0.992 | 0.992 | 0.522 | 0.583 | 0.546 | 0.588 |

Note: [1] the posterior probabilities were calculated using an SVM classifier. Here, the cutoff for q-value in SAM-GS part is set at 0.05.

**Table 2. Performance of both simple SAMGSR and two-level SAMGSR on simulated data**

| Method | | | Time 1 | Time 2 | Time 3 | Time 4 | Time 5 |
|---|---|---|---|---|---|---|---|
| A. Simulation 1 | | | | | | | |
| Simple SAMGSR (Ave. # 32.06) | | # of genes | 19.84 | 19.14 | 13.68 | 9.30 | 11.00 |
| | | F13A1 (%) | 72 | 100 | 100 | 92 | 68 |
| | | GSTM1 (%) | 0 | 0 | 62 | 22 | 0 |
| Two-level SAMGSR (C2: Ave. # 61.74) | | # of genes | 38.88 | 32.66 | 21.44 | 18.96 | 20.50 |
| | | F13A1 (%) | 64 | 92 | 90 | 84 | 52 |
| | | GSTM1 (%) | 2 | 62 | 94 | 80 | 36 |
| B. Simulation 2 | | | | | | | |
| Simple SAMGSR | | # of genes | 182.38 | 56.18 | 35.44 | 30.94 | 123.84 |
| | | COX4I2 (%) | 96 | 0 | 0 | 0 | 4 |

| (Ave. # 291.98) | RP9 (%) | 10 | 4 | 4 | 6 | 96 |
|---|---|---|---|---|---|---|
| Two-level SAMGSR (Ave. # 327.56) | # of genes | 209.44 | 73.40 | 48.04 | 49.38 | 138.66 |
| | COX4I2 (%) | 100 | 0 | 0 | 0 | 0 |
| | RP9 (%) | 4 | 0 | 0 | 0 | 92 |

# Figures

**Figure 1. Flowcharts illustrating both simple and two-level SAMGSR algorithms.**

A. Simple SAMGSR

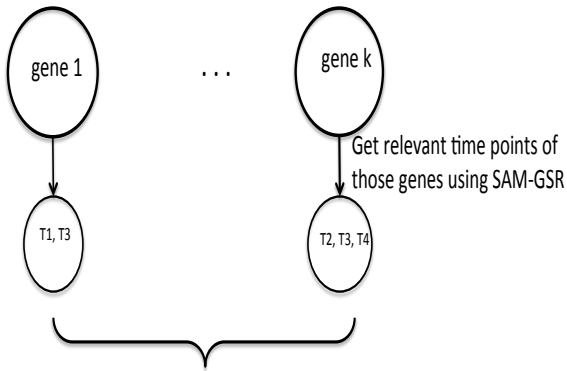

B. Two-level SAMGSR

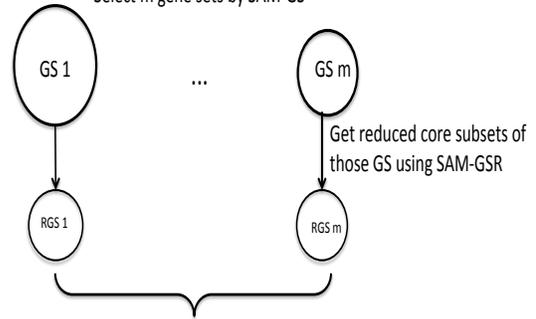

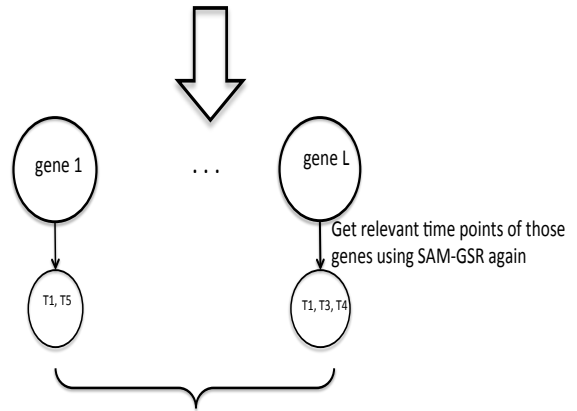

**Figure 2. Characteristics of selected genes by both SAMGSR extensions in the traumatic injury application.** A. Venn-diagram illustrates how the selected genes by simple SAMGSR method overlap at different time points. B. Venn-diagram illustrates how the selected genes by two-level SAMGSR method overlap at different time points. C. Venn-diagram illustrates how the concordantly differentially expressed genes across all time points by both methods overlap.

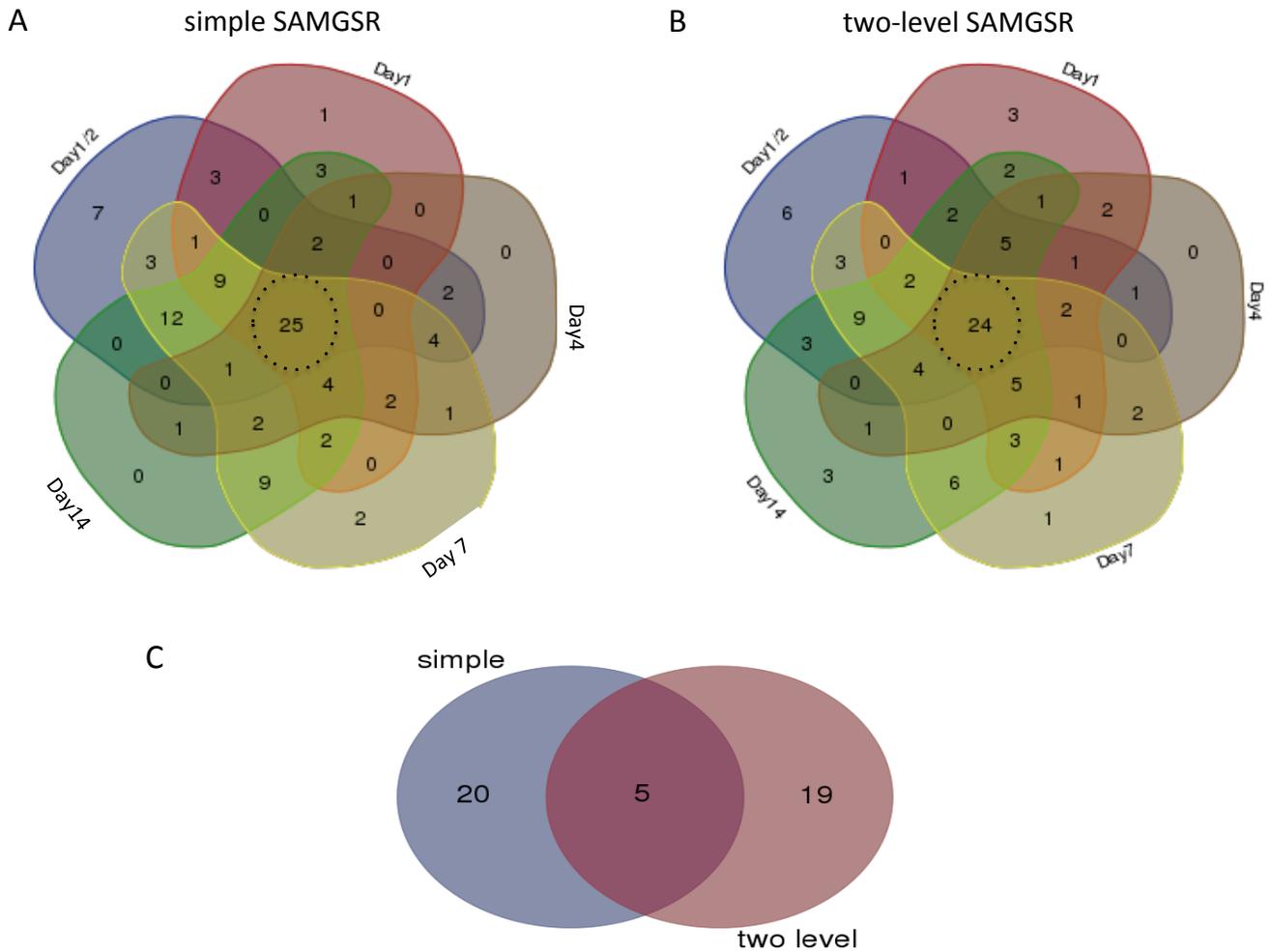

**Figure 3. Characteristics of 5 common significant expressed genes across all time points by both two-level SAMGSR and simple SAMGSR in the traumatic injury application.** Subgroup sample means versus time plot for the 5 common genes that were identified as significance across all 5-time points between uncomplicated and complicated patients. Red line is for complicated group while black line is for uncomplicated group.

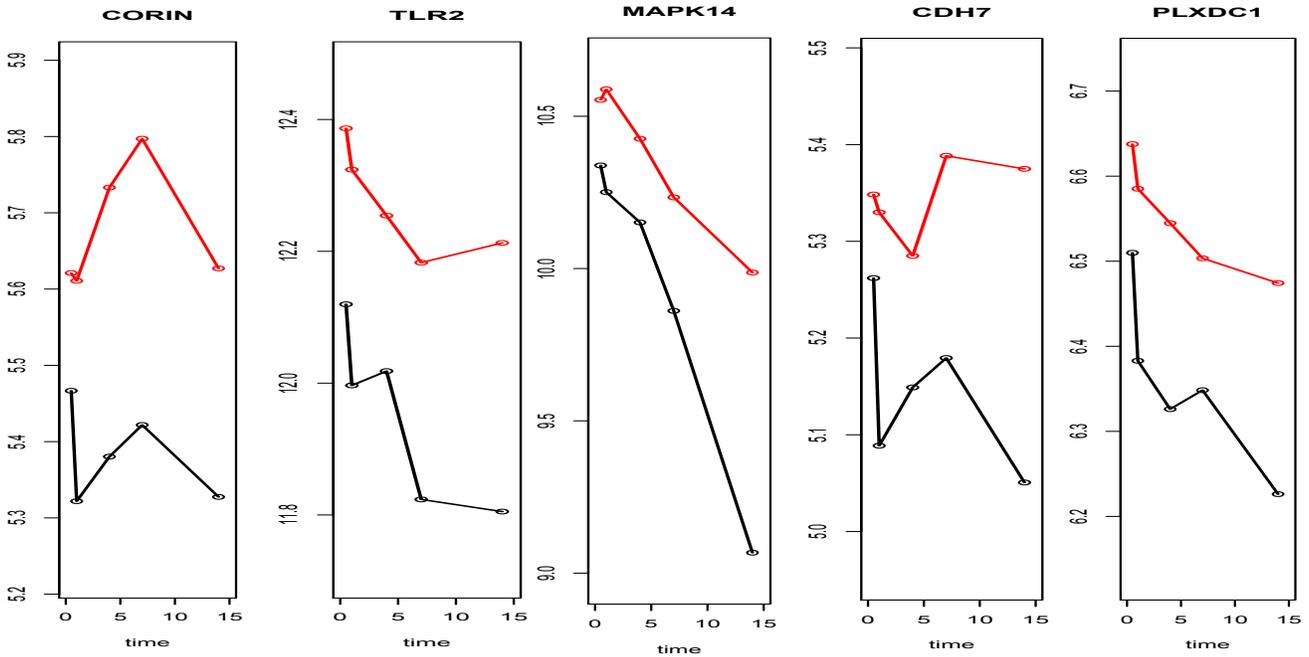